\documentclass{ws-procs9x6}

\begin{document}

\title{Large N gauge theories -- Numerical results}

\author{R. Narayanan}

\address{Department of Physics, Florida International University,\\
Miami, FL 33199, USA \\
$^*$E-mail: rajamani.narayanan@fiu.edu}

\author{ H. Neuberger}

\address{Rutgers University, Department of Physics and Astronomy,\\
Piscataway, NJ 08855, USA\\
$^*$E-mail: neuberg@physics.rutgers.edu }

\begin{abstract} Some physical results in four dimensional
large N gauge theories on a periodic torus
are summarized.
\end{abstract}


\bodymatter

\section{Introduction}\label{sec1}
The large N limit of
four dimensional non-abelian gauge theories 
is interesting from the view point of 
QCD phenomenology~\cite{Manohar:1998xv}
and string theory~\cite{Aharony:1999ti}.
Lattice QCD is a useful technique for extracting
fundamental results in the large N limit of QCD.
Fermions are naturally quenched in the 't Hooft
limit of large N QCD and this significantly
reduces the computational cost in a lattice
calculation. In addition, there is a concept
of continuum reduction~\cite{Kiskis:2003rd}, namely,
physics
does not depend on the size of box $l^4$
for $l > l_c$ and $l_c$ is a physical critical size.
These two observations have been used to extract
physical results in the large N limit of QCD using
numerical techniques on the lattice.

\section{Phases of large N QCD}
  
\begin{figure}
\vskip 1cm
\begin{center}
\psfig{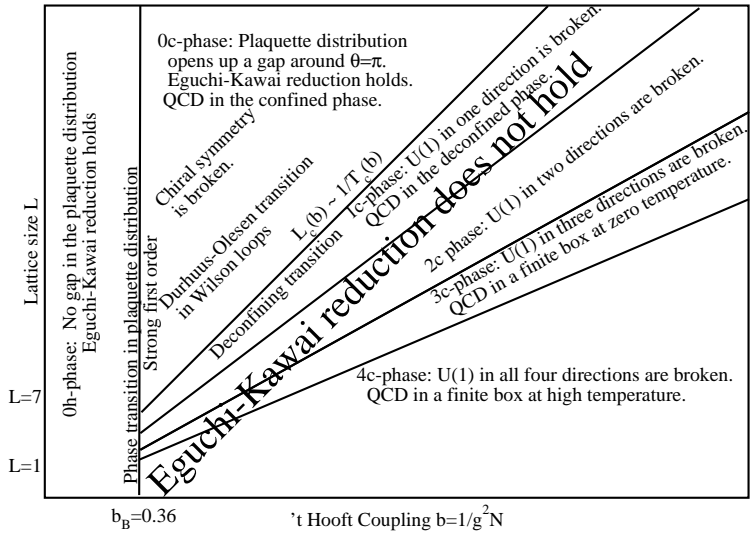}
\end{center}
\caption{The various phases of
four dimensional large N QCD as viewed from the lattice.
}
\label{fourd}
\end{figure}

Large N QCD on a continuum torus $l^4$ has several
phases~\cite{Narayanan:2005en} depending upon the size of the torus
as shown in Fig.~\ref{fourd}. The continuum action has
$U^4(1)$ symmetries associated with the Polyakov loops in the
four directions and the various phases correspond to the
number of directions in which this symmetry is broken.
The continuum limit is obtained by going to the top-right
corner of Fig.~\ref{fourd} and different approaches to this
corner will result in one of the five continuum phases.

The $0$h-phase present for $b < 0.36$ for all $L$ is an
unphysical phase that does not survive the continuum limit.
The $0$h to $0$c transition is associated with the single
plaquette operator opening up a gap around $\pi$ in its eigenvalue
distribution. Gauge fields come in disconnected pieces
in all the $X$c-phases due to the presence of the gap in
the single plaquette operator~\cite{Kiskis:2002gr}.

The $0$c-phase is the confined phase of large N QCD and
the $1$c-phase is the deconfined phase
and $l_c=1/t_c$. An immediate
consequence of continuum reduction is that large N QCD
does not feel temperatures below $t_c$~\cite{Cohen:2004cd}.
Numerical analysis has shown that lattice spacing effects
are small in the $0$c-phase and it is sufficient to work at
$L\sim 9$ and $N\sim 30$ to extract continuum results. Therefore,
numerical computations can be performed on a serial computer
and a cluster of computers can be efficiently used to
generate statistics in a Monte-Carlo calculation.

\section{Chiral symmetry breaking in finite 
volume}

\begin{figure}
\vskip 1cm
\begin{center}
\psfig{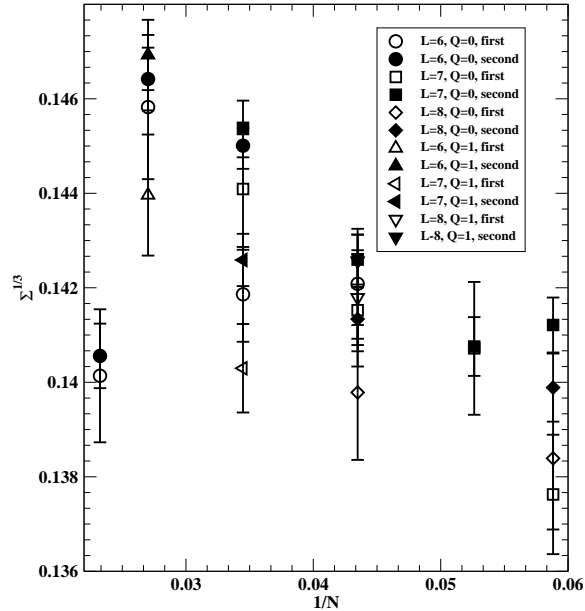}
\end{center}
\caption{Extraction of the chiral condensate in large N
lattice QCD at a fixed lattice coupling.
}
\label{sigma}
\end{figure}

Since physics does not depend on the box size in $0$c-phase,
one should show that chiral symmetry is spontaneously broken
in finite volume in order to properly reproduce physics in
this phase~\cite{Narayanan:2004cp}. 
The order of limits are important and one has
to take the large N limit before taking the quark mass to
zero at finite phsyical volume. The low lying spectrum of the 
massless Dirac operator shows evidence for spontaneous
chiral breaking since the eigenvalues, $i\lambda$,
scales like $ z = \lambda \Sigma N l^4 $ with $z$
obeying a universal distribution. The chiral
condesate, $\Sigma$, is independent of $l$. Results of a 
calculation of the chiral condensate on the lattice is
shown in Fig.~\ref{sigma}. Results from chiral random matrix
theory~\cite{Verbaarschot:2000dy} were used to extract the
chiral condensate at a fixed $N$, $L$ and lattice coupling.
Two lowest non-zero eigenvalues in the $Q=0$ and $Q=1$
topological sectors were used to show consistency. 
The plot shows that there is a limit
as $N\rightarrow\infty$ and this limit is independent of $L$.
Results obtained at different lattice couplings yield
$\frac{l_c^3}{N}\lbrace \bar\psi\psi \rbrace \approx (0.65)^3.$

\section{Pions in large N QCD}

\begin{figure}
\vskip 1cm
\begin{center}
\psfig{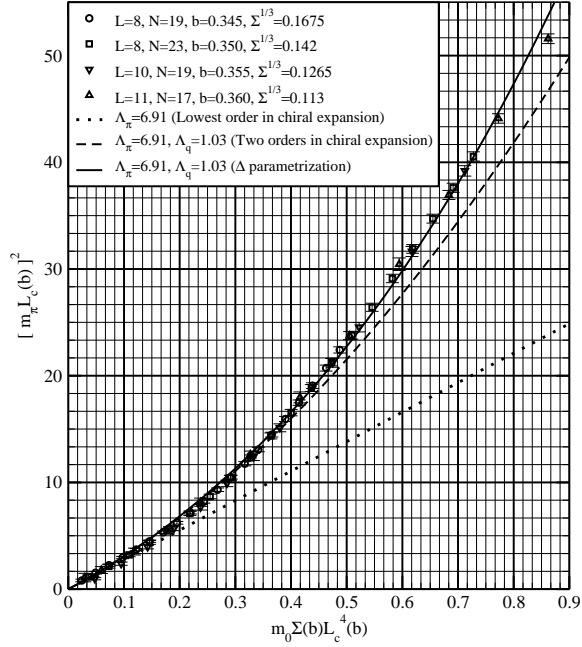}
\end{center}
\caption{The  pion mass as a function of quark mass
in large N QCD.
}
\label{pion}
\end{figure}

Since chiral symmetry is broken in large N QCD even in finite
volume, one should be able to observe massless pions in
finite volume. This result emerges in the 
following manner~\cite{Narayanan:2005gh}.
Properties of a single quark in a background gauge field
$A_\mu(x)$ cannot depend on a shift of
$A_\mu(x) \rightarrow A_\mu(x) + \frac{p_\mu}{l}$ for arbitrary
$p_\mu$ since the $U^4(1)$ symmetries associated with
the Polyakov loops are not broken in the $0$c-phase.
But the propagator of a non-singlet meson will depend
on $p_\mu$, if one quark sees 
$A_\mu(x) \rightarrow A_\mu(x) + \frac{p_\mu}{2l}$ 
and the other sees
$A_\mu(x) \rightarrow A_\mu(x) - \frac{p_\mu}{2l}$ 
as their respective gauge fields. This is referred
to as the quenched momentum prescription~\cite{Gross:1982at} for the 
computation of meson propagators in the large N limit
of QCD. One can use the results for the chiral condensate, $\Sigma(b)$,
and critical lattice size, $L_c(b)$, to plot the pion mass
as a function of $m<\bar\psi\psi> $. The results fall on
a single universal curve as shown in
fig.~\ref{pion} and 
$f_\pi l_c= \frac{1}{\sqrt{2\Lambda_\pi}} \approx 0.269$.

\section{Chiral symmetry restoration}

\begin{figure}
\vskip 1cm
\begin{center}
\psfig{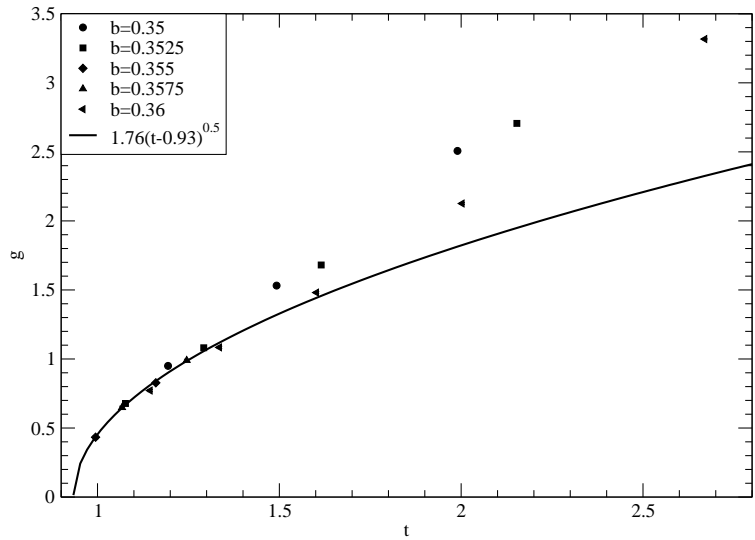}
\end{center}
\caption{The gap in the quark spectrum as a function of 
temperature
in large N QCD.
}
\label{gap}
\end{figure}

The $0$c to $1$c phase transition is the confinement-deconfinement
phase transition since the $U(1)$ symmetry associated
with one of the Polyakov loops is broken. This transition is
first order since there is a latent heat associated with
the single plaquette~\cite{Kiskis:2005hf}. The fermion determinant
does matter in the $1$c-phase and it picks the correct
boundary conditions for fermions in the broken direction,
namely, anti-periodic with 
respect to the Polyakov loop~\cite{Narayanan:2006sd}.
The lowest eigenvalue of the Dirac operator with the
correct boundary conditions can be used to study the gap
in the $1$c-phase and one finds that chiral symmetry
is restored in the $1$c-phase~\cite{Narayanan:2006sd}. 
Furthermore, the 
chiral transition is first order as shown
in Fig.~\ref{gap}. If one were to super-cool the $1$c-phase
into the $0$c-phase, a second order transition with a 
square root singularity would be observed at $0.93/l_c$.

\section{Phase transition in the Wilson loop operator}

Non-abelian gauge theories in the confined phase are
strongly interacting at large distances and weakly
interacting at short distances. If this is seen as
a phase transition in some observable, one could
use the universal behavior of this transition to
connect the low energy physics of QCD to the high 
energy physics of QCD. The Wilson loop operator
as a function of its size is the most likely
candidate to study this transition within the $0$c-phase
of large N QCD~\cite{Narayanan:2006rf}. Such a
transition exists in two dimensional large N QCD~\cite{Durhuus:1980nb}
and it is claimed that the transition in four
dimensional QCD is in the same universality class.
This transition is referred to as the Durhuus-Olesen
phase transition in Fig~\ref{fourd}.

\begin{figure}
\vskip 1cm
\begin{center}
\psfig{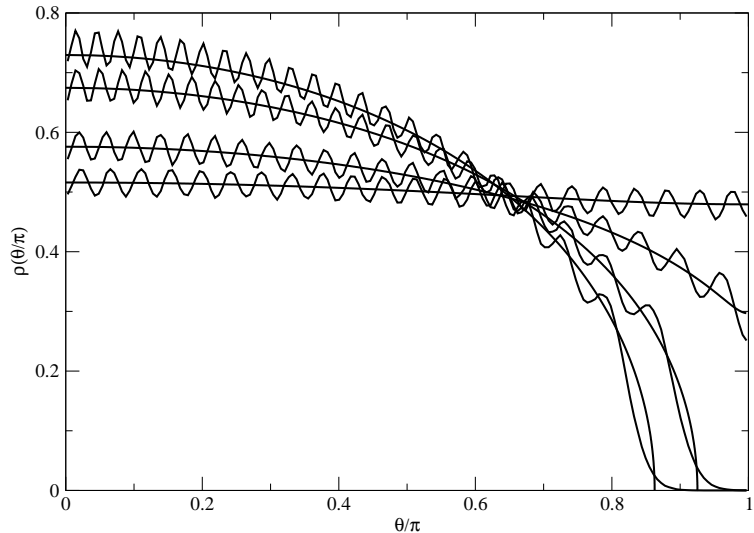}
\end{center}
\caption{The phase transition in the expectation value
of the Wilson loop operator as a
function of its size in large N QCD. The distribution
of the eigenvalues of the smeared Wilson loop operator
are compared to the Durhuus-Olesen distributions.
Results are shown for loops of sizes $l/l_c=0.740,0.660,
0.560,0.503$. They match with $k=4.03,2.30,1.41,1.15$
where $k$ is the dimensionless area in two dimension large N
QCD. $k=2$ is the critical area.
}
\label{wilson}
\end{figure}

Wilson loop operators suffer from a perimeter divergence
in four dimensions and one has to eliminate it when defining
this operator if one were to study its
eigenvalue distribution. One possible way to
suppress the perimeter divergence is to use smeared
operators on the lattice. Numerical studies of the
eigenvalue distribution of the smeared Wilson loop
operator on the lattice shows clear evidence for
a phase transition as a function of the size of
the Wilson loop~\cite{Narayanan:2006rf}. 
The universality class is the same
as the one found in two dimensional large N QCD
as shown in Fig.~\ref{wilson} and the critical
loop size is roughly $0.6 l_c$.

\section*{Acknowledgements}
R. N. would like to thank the
organizers of CAQCD-06 for a stimulating atmosphere.
R. N. acknowledges partial support by the NSF under
grant number PHY-0300065 and
partial support from Jefferson
Lab. The Thomas Jefferson National Accelerator Facility
(Jefferson Lab) is operated by the Southeastern Universities Research
Association (SURA) under DOE contract DE-AC05-84ER40150.
H. N. acknowledges partial support
by the DOE under grant number
DE-FG02-01ER41165 at Rutgers University.

\end{document}